\documentclass{PoS}

\title{Computing the long-distance contribution to the kaon mixing 
parameter $\epsilon_K$}

\ShortTitle{Long-distance contribution to $\epsilon_K$}

\author{\speaker{Norman Christ}%
        \thanks{This work was supported in part by U.S. 
        DOE grant DE-FG02-92ER40699}\\
        Department of Physics \\
        Columbia University, USA\\
        E-mail: \email{nhc@phys.columbia.edu}}

\author{RBC and UKQCD collaborations}

\abstract{The largest contribution to the CP violating
$K_L-K_S$ mixing parameter $\epsilon_K$ comes from
second order weak interactions at short distances 
and can be accurately determined by a combination
of electroweak perturbation theory and the 
calculation of the parameter $B_K$ from lattice QCD.  
However, there is an additional long distance 
contribution to $\epsilon_K$ which is estimated to 
be of order 5\%.  Here recently introduced lattice 
techniques for computing the long-distance component 
of the $K_L-K_S$ mass difference are generalized 
to this long-distance contribution to $\epsilon_K$.}

\FullConference{XXIX International Symposium on Lattice Field Theory \\
		 July 10 – 16 2011\\
		 Squaw Valley, Lake Tahoe, California}

\begin{document}

\section{Introduction}

Some of the most stringent tests of the standard model and
most promising windows into possible new physics come from
process which occur at second order in the Fermi constant
$G_F$.  These include the standard model prediction for indirect
CP violation in the $K^0 - \overline{K}^0$ system and the bound
on interactions which change strangeness by two units coming
from the small $K_L-K_S$ mass difference.  Such $O(G_F^2)$
amplitudes involve the exchange of two intermediate bosons 
which requires that each pair of vertices joined by such a
$W^\pm$ exchange are separated by a short distance on the order
of $1/m_W$.  However, the spatial separation between these
two pairs of vertices need not be small, and ``long distance'' 
separations of the order of $1/\Lambda_{\rm QCD}$ are possible.

This is well illustrated by the familiar Wigner-Weisskopf formula 
describing the time evolution of the $K^0 - \overline{K}^0$ system:
\begin{equation}
i\frac{d}{dt}\left(\begin{array}{c} K^0 \\ \overline{K}^0 \end{array}\right) 
= \left\{ \left( \begin{array}{cc} M_{00} & M_{0\overline{0}} \\
                        M_{\overline{0}0} & M_{\overline{0} \overline{0}}
                           \end{array} \right) 
- \frac{i}{2} \left( \begin{array}{cc} \Gamma_{00} & \Gamma_{0\overline{0}} \\
               \Gamma_{\overline{0}0} &\Gamma_{\overline{0} \overline{0}}
                           \end{array} \right)\right\} 
\left(\begin{array}{c} K^0 \\ \overline{K}^0 \end{array}\right)
\label{eq:WW}
\end{equation}
where the $2 \times 2$ matrices $M$ and $\Gamma$ are given by:
\begin{eqnarray}
\Gamma_{ij} &=& 2\pi \sum_\alpha \int_{2m_\pi}^\infty d E
                      \langle i |H_W|\alpha(E)\rangle
                      \langle \alpha(E)|H_W|j\rangle \delta(E-m_K)  
\label{eq:WW-Gamma} \\
M_{ij} &=& \sum_\alpha {\cal P} \int_{2m_\pi}^\infty d E \frac{
                       \langle i |H_W|\alpha(E)\rangle
                       \langle \alpha(E)|H_W|j\rangle}{m_K - E}. 
\label{eq:WW-M}
\end{eqnarray}
We are using the subscripts $0$ and $\overline{0}$ to represent the 
$K^0$ and $\overline{K}^0$ states.  The generalized sum over $\alpha$ 
and integral over the energy represents the sum over a complete set 
of energy eigenstates normalized as $\langle\alpha'(E')|\alpha(E)\rangle
=\delta(E'-E)\delta_{\alpha'\alpha}$.  These two matrices determine the 
two important quantities, $m_{K_S}-m_{K_L}$ and $\epsilon_K$:
\begin{eqnarray}
m_{K_S}-m_{K_L} &=& 2 {\rm Re}\{M_{0\overline{0}}\}
\label{eq:delta_m_K-1} \\
\epsilon_K &=& \frac{i}{2}\left\{\frac{{\rm Im} M_{0\overline{0}}
                            -\frac{i}{2}{\rm Im} \Gamma_{0\overline{0}} }
                                  {{\rm Re} M_{0\overline{0}}
                            -\frac{i}{2}{\rm Re} \Gamma_{0\overline{0}} }
\right\}
                 + i\frac{{\rm Im}A_0}{{\rm Re}A_0} \label{eq:epsilon_K}
\end{eqnarray}

The absorptive part, given by the energy conserving matrix $\Gamma_{ij}$,
is formed from products of first-order weak amplitudes which are now the 
targets of increasingly precise lattice calculations, see for example 
Refs.~\cite{Blum:2011pu} and \cite{Blum:2011zz}.  The dispersive part, 
$M_{ij}$ is intrinsically second order in $G_F$ and contains both a 
short-distance part were both exchanged $W$'s are separated by distance 
scales much smaller than $1/\Lambda_{\rm QCD}$ and the long-distance part 
discussed above.  

The computation of the short distance contributions to $\epsilon_K$ or 
$m_{K_S}-m_{K_L}$ is now a highly successful application of lattice QCD.
At the scale of QCD, phenomena taking place at a distance scale of 
$1/m_W$ can be accurately represented by an effective four-Fermi 
operator whose coefficient can be computed in electro-weak perturbation
theory and whose matrix element between kaon states can be computed using
lattice QCD.  For $\epsilon_K$ these short-distances dominate and the 
long-distance part is estimated to be a few percent 
correction~\cite{Buras:2010pza}.  However, current results for this 
short contribution to $\epsilon_K$ are now accurate on the 5\% level, 
giving the unknown long-distance part increasing importance.  For the 
$K_L-K_S$ mass difference the size of the long-distance contribution is 
less certain and could provide as much as 50\% of the full result.  

It is the lattice calculation of these long distance parts which 
is the subject of this talk.  We will briefly review the approach 
presented in Ref.~\cite{Christ:2010zz} for the calculation of the 
long-distance contribution to the $K_L-K_S$ mass difference and 
then describe its generalization to the more complex case of 
$\epsilon_K$.

\section{Strategy for lattice calculation $K_L-K_S$ mass difference.}

As presented in Ref.~\cite{Christ:2010zz} a Euclidean-space, finite
volume calculation of the $K_L-K_S$ mass difference $\Delta m_K$ contains
three important ingredients.  The first is a relation between the second-order, 
infinite-volume, principal part integral in Eq.~\ref{eq:WW-M} giving 
$\Delta m_K = 2{\rm Re}(M_{\overline{0}0})$ and a finite-volume, discrete 
perturbation theory sum.  This relation is obtained through a generalization 
of the first-order formula of Lellouch and Luscher for the $K\to\pi\pi$ 
decay width~\cite{Lellouch:2000pv}.  The second ingredient is a computational 
strategy for evaluating that perturbation theory sum using lattice QCD.  
While with sufficient statistics the resulting lattice calculation will 
correctly determine the long-distance part of $\Delta m_K$, the short 
distance contribution to this result will be incorrect, reflecting the 
details of the lattice regulator.  Thus, the third ingredient is a method 
to replace this erroneous short-distance contribution with the correct 
short distance part of the continuum theory.  We will now briefly review 
each of these in turn.

The relation between $\Delta m_K$ and a finite volume 
perturbation theory sum follows from the Luscher finite-volume quantization 
condition~\cite{Luscher:1990ux} relating the allowed finite volume energies, 
$E_n$, for a two-pion system and the $\pi-\pi$ scatting phase shift:
\begin{equation}
\phi(k_n L/2\pi) + \delta_0(E_n) + \delta_W(E_n) = n\pi
\label{eq:Luscher}
\end{equation}
where $k_n=\sqrt{E_n^2/4-m_\pi^2}$ and the known function $\phi(y)$ is 
defined in Ref.~\cite{Luscher:1990ux}.  Here we have divided the $\pi-\pi$ 
scatting phase shift into the strong $s$-wave phase shift $\delta_0$ and
that caused by the resonant scattering through the weakly coupled $K_S$
state.  Since the location of the kaon pole in $\delta_W(E_n)$ is shifted 
by $\Delta m_K$, Eq.~\ref{eq:Luscher} can be used to relate $\Delta m_K$
to the finite-volume, second-order perturbation theory sum which determines
$E_n$.  It is this argument which we will generalize to determine 
Im$(M_{\overline{0}0})$.

Since the resulting finite volume sum that must be evaluated to determine
$\Delta m_K$ is closely related to perturbation theory, the entire sum
can be obtained from an integrated Green's function constructed from four
operators: two interpolating operators which create an initial $K^0$ state 
at a time $t_i$ and destroy a final $\overline{K}^0$ state at $t_f$ and 
two $\Delta S=1$ weak operators $H_W(t_1)$ and $H_W(t_2)$ evaluated at 
intermediate times and integrated over a range 
$t_f \gg t_b \ge t_k \ge t_a \gg t_i$ for $k=1$, 2.  The term 
proportional to length of the integration interval, $t_b-t_a$ gives the 
desired perturbation theory sum which determines $\Delta m_K$.  While 
the evaluation of this time-integrated four-point function poses
substantial computational challenges, our first exploratory study 
gives encouraging results~\cite{Yu:2011gk}.

Since $\Delta m_K$ receives both short- and long-distance contributions,
the integrated amplitude described in the previous paragraph will receive
important contributions from the region in which the two weak vertices 
coincide.  If a charm quark is included to exploit GIM cancelation a
substantial, incorrect short-distance contribution will remain, of size 
$\ln(m_W a)$.  Fortunately, the same techniques which allow 
this continuum short-distance part to be precisely defined and evaluated 
in a lattice calculation can be employed here to isolate and replace 
this lattice artifact by the correct continuum piece.  If the four quarks
making up the two kaon interpolating operators in the above four-point
function are instead treated separately and the resulting Green's function
evaluated in Landau gauge and given large, off-shell momentum the
short distance contribution can be evaluated numerically in a fashion
that is free of infrared singularity allowing a simple substitution by
the correct continuum short distance part.  Again, first numerical 
experiments~\cite{Yu:2011gk} suggest that this can be done without 
difficulty.

\section{Extension to $\epsilon_K$} 

We now turn to the central topic of this talk, the extension of the method
reviewed above to the case of Im$(M_{\overline{0}0})$ and $\epsilon_K$.
Since there are no apparent added difficulties associated with the
numerical evaluation of the resulting finite-volume expression or the
correction of its short-distance part these topics will not be discussed
further.  However, on first-sight using a Lellouch-Luscher style argument
to obtain a finite volume formula for Im$(M_{0\overline{0}})$ may seem
unlikely to succeed.  First $\epsilon_K$ describes the mixing of the $K_S$
and $K_L$ states and the $K_L$ state with its predominate decay into three
pions would appear inaccessible to Luscher's formalism which applies only
to two-particle scattering.  Further the energies that are the basis of
the Lellouch-Luscher method are intrinsically CP conserving suggesting
no sensitivity to the CP violating phase of $M_{0\overline{0}}$.  

Fortunately both of these problems can be avoided by adding a fictitious, 
$\Delta S=2$ ``superweak'' term to the Hamiltonian whose effects can be 
combined with (but will not alter at leading order) those of the standard 
weak interactions:
\begin{equation}
H^{\rm SW} = (\omega_r +i\omega_i)\left\{\overline{s}(1+\gamma^5)\gamma^\mu d \;
                                     \overline{s}(1+\gamma^5)\gamma^\mu d\right\}
                                     +\mbox{hermitian conjugate}.
\label{eq:SW}
\end{equation}
The quantities $\omega_r$ and $\omega_i$ are then chosen so that
\begin{eqnarray}
\omega_r\langle \overline{K}^0|O_{LL}|K^0\rangle+{\rm Re}M_{\overline{0}0} &=& 0
\\
\omega_i\langle \overline{K}^0|O_{LL}|K^0\rangle \gg |\Gamma_{ij}|,
\end{eqnarray}
for all $i$ and $j$.  Here $O_{LL}$ is the operator within the curly 
brackets in Eq.~\ref{eq:SW}.  With the off-diagonal, real part of $M_{ij}$
canceled and the off-diagonal term proportional to $\omega_i$ dominating
the off-diagonal parts of $\Gamma_{ij}$, the eigenstates of the time
development operator in Eq.~\ref{eq:WW} become the unfamiliar combinations
$K_\pm = (|K^0\rangle \pm i|\overline{K}^0\rangle)/\sqrt{2}$ --- both states
which decay predominantly into two pions, consistent with Luscher's finite
volume formalism.  In addition, the large contribution of $\omega_i$ insures
that the two $\pi-\pi$ resonances associated with the states $K_\pm$ are
non-overlapping and will each contribute independent resonant behavior to 
that scattering.  The eigenvalues of the $2 \times 2$ time development 
matrix on the right-hand side of Eq.~\ref{eq:WW} for the states 
$|K_\pm\rangle$ can be written:
\begin{eqnarray}
\lambda_\pm &=& \frac{1}{2}\left(M_{00}+M_{\overline{0}\overline{0}}\right)
      \mp\left( \omega_i \langle K^0|O_{LL}|\overline{K}^0\rangle
    +  {\rm Im}M_{0\overline{0}}\right) \nonumber \\
  &&\hskip 1.0in     + i\frac{1}{4}\left(\Gamma_{00}+\Gamma_{\overline{0}\overline{0}}
                              \mp i\Gamma_{\overline{0}0} 
                              \pm i\Gamma_{0\overline{0}} \right) \\
&=& M_{00} \mp\left( \omega_i \langle K^0|O_{LL}|\overline{K}^0\rangle
    +  {\rm Im}M_{0\overline{0}}\right)
    + i\frac{1}{2} \left(\Gamma_{00}\mp {\rm Im}\Gamma_{0\overline{0}} \right)
\label{eq:eigenvalues_iv}
\end{eqnarray}

We first examine the finite volume problem.  Here the box size $L$ is 
adjusted so that we have three nearly degenerate states: the 
$\pi-\pi$ state $|n_0\rangle$ with energy $E_{n_0}$ as well as the 
two single-particle states, $|K^0\rangle$ and $|\overline{K}^0\rangle$.  
Using second order, degenerate perturbation theory, we can determine 
the finite volume energies of interest as the eigenvalues of the 
$3\times 3$ matrix:
\begin{equation} 
\left(\begin{array}{ccc}
m_K + {\sum \atop {n \ne n_0}}\frac{|\langle n|H_W|K^0\rangle|^2}{m_K-E_n}
            & {\sum \atop {n \ne n_0}}\frac{\langle K^0|H_W|n\rangle
                         \langle n|H_W|\overline{K}^0\rangle}{m_K-E_n}
              +\delta m
                            & \langle K^0|H_W|n_0\rangle            \\
{\sum \atop {n \ne n_0}}\frac{\langle \overline{K}^0|H_W|n\rangle
                         \langle n|H_W|K^0\rangle}{m_K-E_n} +\delta m^*
            & m_K 
   + {\sum \atop {n \ne n_0}}\frac{|\langle n|H_W|\overline{K}^0\rangle|^2}
               {m_K-E_n}
                            & \langle \overline{K}^0|H_W|n_0\rangle \\
\langle n_0|H_W|K^0\rangle 
            & \langle n_0|H_W|\overline{K}^0\rangle              
                            & E_{n_0} + {\sum \atop {n \ne K^0,\overline{K}^0}}
                     \frac{|\langle n|H_W|n_0\rangle|^2}{E_{n_0}-E_n}
\end{array}\right),
\label{eq:3x3}
\end{equation}
with the rows and columns corresponding to $|K^0\rangle$, 
$|\overline{K}^0\rangle$ and $|n_0\rangle$ in increasing order.
Here the complex quantity $\delta m$ is given by:
\begin{equation}
\delta m = (\omega_r+i\omega_i)\langle K^0|O_{LL}|\overline{K}^0\rangle. 
\end{equation}

Using the CP and time-reversal operations, the $3 \times 3$ matrix in
Eq.~\ref{eq:3x3} can be written as:
\begin{equation}
\left(\begin{array}{ccc}
E_{d1}                     & {\cal M}           & e^{-i\delta_0}{\cal A}   \\
{\cal M}^*                 & E_{d1}             & e^{-i\delta_0}{\cal A}^*   \\
e^{i\delta_0}{\cal A}^*    & e^{i\delta_0}{\cal A}
                                                & E_{d2}
\end{array}\right).
\label{eq:3x3_simple}
\end{equation}
The eigenvalues of this matrix can be determined by diagonalizing the 
large part of the matrix proportional to the first-order amplitude
${\cal A}$ and then applying perturbation theory to diagonalize
the remainder to order $H_W^2$:
\begin{eqnarray}
E_0   &=& E_{d1}-\frac{{\rm Re}\left\{{\cal M}({\cal A}^*)^2\right\}}
                    {|{\cal A}|^2} \\
E_\pm &=& \frac{1}{2}\left(E_{d1}+E_{d2}\right) \pm\sqrt{2}|{\cal A}|
          +\frac{{\rm Re}\left\{{\cal M}({\cal A}^*)^2\right\}}
                    {2|{\cal A}|^2}.
\label{eq:eigenvalue_fv_pm}
\end{eqnarray}

We now require that the Luscher finite volume condition, 
Eq.~\ref{eq:Luscher}, be obeyed when the total $\pi-\pi$ 
phase shift, written as a combination of the usual strong 
$s$-wave phase shift, the resonant scattering through the 
$|K_\pm\rangle$ states and a standard second-order-weak 
Born term is evaluated at the finite-volume energy
eigenvalues $E_\pm$ given in Eq.~\ref{eq:eigenvalue_fv_pm}: 
\begin{eqnarray}
n\pi &=& \phi\left(\frac{k_\pm L}{2\pi}\right) + \delta_0(E_\pm) 
   + \arctan\left(\frac{\Gamma_+(E_\pm)/2}{M_+ -E_\pm}\right) 
   + \arctan\left(\frac{\Gamma_-(E_\pm)/2}{M_- -E_\pm}\right)
\label{eq:LL-generalized} \\
   &&\hskip 2.0in    -\pi\sum_{\beta \ne K^0,\overline{K}^0} 
              \frac{|\langle \beta|H_W|\pi\pi(E)\rangle|^2}{E -E_\beta}.
\nonumber
\end{eqnarray}
Here $M_\pm$ and $\Gamma_\pm$ can be determined from the real and 
imaginary parts of the infinite-volume eigenvalues given in 
Eq.~\ref{eq:eigenvalues_iv}:
\begin{eqnarray}
M_\pm      &=& m_K + M_{00} 
         \mp\left( \omega_i \langle K^0|O_{LL}|\overline{K}^0\rangle
    +  {\rm Im}M_{0\overline{0}}\right)  \\
\Gamma_\pm &=& \Gamma_{00}\mp {\rm Im}\Gamma_{0\overline{0}}\,.
\end{eqnarray}

If the finite volume is adjusted so that the zero-order $\pi-\pi$
energy $E_{n_0} = m_K$, then to zeroth order in $H_W$, 
Eq.~\ref{eq:LL-generalized} reduces to the original Luscher quantization 
condition relating the strong $s$-wave phase shift $\delta_0(E_\pm)$ and 
the finite-volume $\pi-\pi$ energy $E_{n_0}$.  If, for the same value
of $E_{n_0}$, Eq.~\ref{eq:LL-generalized} is expanded to first order in
$H_W$ then the Lellouch-Luscher relation between the finite volume
$K-\pi\pi$ matrix element $\cal A$ and the $K\to\pi\pi$ width 
$\Gamma_{00}$ is obtained.

Considerably more algebra is needed to obtain a relation 
between the finite volume sum in $\cal M$ and the infinite volume
amplitude $M_{0\overline{0}}$.  As in the earlier study of 
$\Delta m_K$, we first evaluate Eq.~\ref{eq:LL-generalized} for a
volume which makes the $\pi-\pi$ energy $E_{n_0}$
close to $m_K$ on the scale of QCD but sufficiently different that
a perturbative expansion in $H_W/(E_{n_0}-m_K)$ is possible.  
Equation.~\ref{eq:LL-generalized} can then be evaluated in
a power series in $H_W$.  The resulting formula contains a pole at
$E_{n_0}=m_K$.  Requiring that this pole term vanish gives
the standard Lellouch-Luscher relation.  Requiring that the non-pole
piece vanishes at $E_{n_0}=m_K$ then relates the infinite-volume,
second-order Born term in Eq.~\ref{eq:LL-generalized} and the finite-volume,
second-order sum in the lower left corner of the matrix
in Eq.~\ref{eq:3x3}.

In the next step, Eq.~\ref{eq:LL-generalized} is evaluated for a
volume making the two-pion energy $E_{n_0}=m_K$ and the result 
expanded to second order in $H_W$, including the resonant denominators.  
The expression that results contains denominators of order $H_W^2$ and 
numerators of order $H_W^4$ and can be written:
\begin{eqnarray}
0 &=& \frac{\partial(\phi+\delta_0)}{\partial E}
         \Biggl\{\sum_{n \ne n_0}\frac{|\langle n|H_W|K^0\rangle|^2}{m_K-E_n}
         -M_{00} \nonumber \\
&&  + \frac{1}{|{\cal A}|^2} {\rm Re}\Biggl( ({\cal A}^*)^2
  \Bigl\{\sum_{n\ne n_0}\frac{\langle K^0|H_W|n\rangle
              \langle n|H_W|\overline{K}^0\rangle}{m_K-E_n} 
  - M_{0\overline{0}} \Bigr\} \Biggr) \Biggr\}
\nonumber \\
  &&     +\frac{\partial^2(\phi+\delta_0)}{\partial E^2}|{\cal A}|^2
 -2\frac{\partial}{\partial E_{n_0}}\left\{\frac{\partial(\phi+\delta_0)}{\partial E}
  |\langle n_0|H_W|K^0\rangle|^2 \right\},
\label{eq:LL2_1}
\end{eqnarray}
giving a finite volume expression for a combination of the infinite 
volume amplitudes $M_{00}$ and $M_{0\overline{0}}$.  

To determine $M_{0\overline{0}}$ alone, a second equation is needed.
As in the simpler case examined Ref.~\cite{Christ:2010zz}, a second
relation follows is we require that the expectation value of the matrix 
$M_{ij}$ for the state $|\widetilde{K}\rangle = 
({\cal A}|K^0\rangle - {\cal A}^*|\overline{K}^0\rangle)/(2|{\cal A}|^2)$
agree with the corresponding finite volume expression up to terms
exponentially small in the size of that volume.  Here the state 
$|\widetilde{K}\rangle$ is chosen so that it does not couple to the
two-pion, on-shell state, justifying the neglect of finite volume effects.
The resulting equation can be combined with Eq.~\ref{eq:LL2_1} to eliminate
$M_{00}$ giving the relation:
\begin{eqnarray}
0 &=& {\rm Re}\Biggl\{ ({\cal A}^*)^2
  \Biggl(\sum_{n\ne n_0}\frac{\langle K^0|H_W|n\rangle
              \langle n|H_W|\overline{K}^0\rangle}{m_K-E_n} 
  - M_{0\overline{0}} 
\label{eq:LL2_2} \\
  &&     + \frac{1}{\frac{\partial(\phi+\delta_0)}{\partial E}}\Biggl[
\frac{1}{2}
\frac{\partial^2(\phi+\delta_0)}{\partial E^2}{\cal A}^2
 -\frac{\partial}{\partial E_{n_0}}\left\{
   \frac{\partial(\phi+\delta_0)}{\partial E}
  \langle K^0|H_W|n_0\rangle\langle n_0|H_W|\overline{K}^0\rangle
 \right\}\Biggr]\Biggr)\Biggr\}.
\nonumber 
\end{eqnarray}
Since the $K\to\pi\pi$ amplitude ${\cal A}$ is non-zero with 
a phase that is independent of the phase of the 
amplitude in curved brackets, $(\ldots)$, we can conclude that 
$(\ldots) = 0$, our desired finite-volume expression for 
$M_{0\overline{0}}$.  The real part of this equation gives
the earlier result for Re$\{M_{0\overline{0}}\}$.  The 
imaginary part gives the new result for Im$\{M_{0\overline{0}}\}$:
\begin{eqnarray}
\mbox{Im}\{M_{0\overline{0}}\} &=& {\rm Im}\Biggl\{ 
  \sum_{n\ne n_0}\frac{\mbox{Im}\left\{\langle K^0|H_W|n\rangle
              \langle n|H_W|\overline{K}^0\rangle\right\}}{m_K-E_n} 
+ \frac{1}{\frac{\partial(\phi+\delta_0)}{\partial E}}\Biggl[
\frac{1}{2}
\frac{\partial^2(\phi+\delta_0)}{\partial E^2}
          \langle K^0|H_W|n_0\rangle\langle n_0|H_W\overline{K}^0\rangle
\nonumber \\
 && \hskip 0.8in -\frac{\partial}{\partial E_{n_0}}\left\{
   \frac{\partial(\phi+\delta_0)}{\partial E}
  \langle K^0|H_W|n_0\rangle\langle n_0|H_W\overline{K}^0\rangle
 \right\}\Biggr]\Biggr\}.
\label{eq:LL2_3}
\end{eqnarray}
Thus, by exploiting the interference between the complex, CP-violating
amplitude ${\cal A}$ and the second-order, Wigner-Weisskopf mass matrix, 
we have been able to determine the CP-violating part of $M_{0\overline{0}}$
from a finite-volume energy which is intrinsically CP-even.

Here the left hand side of Eq.~\ref{eq:LL2_3} is the infinite volume 
quantity which determines the long- and short-distance dispersive 
parts of $\epsilon_K$.  The right-hand side contains only finite volume 
matrix elements which can be evaluated in a lattice QCD calculation.
The similarity of the results for Re$\{M_{0\overline{0}}\}$, Im$\{M_{0\overline{0}}\}$
and $M_{00}$ (not displayed) suggests that a more direct derivation
may be possible.

\section{Conclusion}

Given the control over the singularity associated with the principal 
part and related finite volume effects implied by these results and 
those in Ref.~\cite{Christ:2010zz}, both the $K_L - K_S$ mass difference 
$\Delta m_K$ and the complete, dispersive part of $\epsilon_K$ can in 
principle be computed using lattice methods.  In fact, the first 
exploratory results of Jianglei Yu, presented in Ref.~\cite{Yu:2011gk}, 
suggest that such a lattice calculation may be possible with the next 
generation of high performance computers.  

However, many obstacles must be overcome: (a) This initial 
work suggests that the short-distance parts of $\Delta m_K$ and 
$\epsilon_K$ can be properly treated by a single subtraction,
defined using Rome-Southampton techniques.  Such a single subtraction 
is adequate only if GIM cancelation has been realized in the lattice 
calculation, requiring the inclusion of the charm quark which in turn 
requires a small lattice spacing.  (b) Many more operators and 
contractions must be included in a complete calculation than have been 
attempted in Ref.~\cite{Yu:2011gk}.  (c) These initial calculations 
use a relatively massive, 410 MeV pion.  The successful subtraction of 
an exponentially growing single-pion contribution demonstrated in 
Ref.~\cite{Yu:2011gk} will become more difficult as the 
pion mass is reduced.  Even greater difficulty will result from the 
vacuum subtraction needed when disconnected graphs are included.  We 
believe that these difficulties can be overcome with improved numerical 
methods and the sustained, 100 Tflops capability of the next generation 
machines now being installed.  
The author acknowledges the important contributions of his RBC/UKQCD 
collaborators to this work.

\bibliography{epsilon_LD}
\bibliographystyle{JHEP}

\end{document}